\documentclass[reprint, amsmath, amssymb, aps, prl]{revtex4-2}

\usepackage{graphicx}
\usepackage{dcolumn}
\usepackage{bm}

\usepackage{xcolor}

\begin{document}

\preprint{APS/123-QED}

\title{Exploring novel compact quasi-axisymmetric stellarators}

\author{T. M. Schuett}
\email{tobias.schuett@york.ac.uk}
\affiliation{%
 York Plasma Institute, School of Physics, Engineering and Technology, University of York, Heslington YO10 5DD, United Kingdom
}%

\author{S. A. Henneberg}%
\affiliation{
Max-Planck-Institut für Plasmaphysik, Wendelsteinstr. 1, 17491 Greifswald, Germany
}%

\date{\today}

\begin{abstract}
The new class of compact quasi-axisymmetric stellarators with a wide range of field periods offers the unique potential to combine the advantages of the two leading magnetic confinement fusion devices, tokamaks and stellarators.
Here we present the first numerical optimization of this class which has so far only been obtained analytically. 
Our approach finds significantly improved quasi-axisymmetric equilibria at aspect ratio $< 2.5$, resulting in no losses of alpha particles at reactor volume while also satisfying improved magnetohydrodynamic stability and a self-consistent plasma current.
\end{abstract}

\maketitle

\textit{Introduction}.\textemdash Thanks to their axisymmetry, tokamaks are one of the most successful and at the same time simplest magnetic confinement fusion devices. 
The magnetic field needed for tokamaks can be simply created with planar coils. 
In addition to their early success, many advances have been made in tokamak research over the past decades, e.g. the discovery of improved confinement regimes \cite{Whyte2010}. 
However, there are limitations to tokamak performance such as easily achieving steady state operation \cite{Ono2015}, 
current driven instabilities which cause disruptions, as well as the empirical Greenwald density limit \cite{Greenwald2002} which can limit performance.

The three dimensionally (3D) shaped, related concept of stellarators offers a path to overcome these limitations \cite{Boozer_2008,Ku_Boozer_2009}. The 3D shaping itself generates twist in the magnetic field (rotational transform $\iota$) necessary for confinement \cite{Helander_2014}, and therefore replaces the requirement of the strong plasma current needed in tokamaks. 
As a result, current driven instabilities and therefore disruptions are typically not an issue in stellarators. 
Steady state operation is straightforward to realize since one does not rely on a plasma current creating the rotational transform. And even beta limits from pressure gradient driven instabilities show to be softer in stellarators \cite{Watanabe2005}. 
And finally, since the geometry strongly impacts the physics properties, their 3D nature can be exploited to optimize the physics performance.
However, stellarators typically require non-planar coils to achieve the 3D vacuum field. 
Reducing the coil complexity as well as achieving a performance similar to that of tokamaks requires careful optimization and typically still a large number of different non-planar coil types.
While small 3D magnetic perturbations are commonly used in tokamaks for instability suppression, they are generated by simple coil systems that are not optimized in shape to maintain performance.

In general, due to the broken symmetry of stellarators, the (neoclassical) confinement can be worse than in tokamaks \cite{Helander_2014}. 
Quasi-symmetry (QS) \cite{Boozer1983,NuehrenbergZille1988,Boozer1995} is able to effectively replace axisymmetry. 
In QS fields the magnetic field strength $B = |\mathbf{B}|$ varies in a special coordinate system (Boozer coordinates ($\psi,\theta_B,\varphi_B$), \cite{Boozer1981_BoozerCoordinates}) only in two coordinates $B = B(\psi,M\theta_B-N\varphi_B)$. Here $\psi$ is the toroidal magnetic flux and serves as a radial coordinate, $\theta_B$ and $\varphi_B$ are the poloidal and toroidal Boozer angles, and $M$ and $N$ specify the flavor of QS. 
Both quasi-axisymmetry (QA) ($N = 0$) and quasi-helical symmetry (QH) ($N \neq 0$) are realizable for $M = 1$ \cite{Helander_2014, landreman2022mappingQS}. 
In practice QS can be achieved with high precision throughout the full plasma volume, reducing neoclassical transport to tokamak levels \cite{LandremanPaul2022}. 
In addition to QS, stellarators are typically designed with another discrete toroidal symmetry, specified by the number of field periods ($N_P$) which describes the number of identical toroidal segments such that $B(\psi,\theta,\varphi + 2\pi/N_P) = B(\psi,\theta,\varphi)$ for any choice of magnetic coordinates ($\psi,\theta,\varphi$). 
Larger $N_P$ can reduce the number of unique coils and hinder electron scale turbulent transport \cite{Plunk2019}.

In previous work from optimization \cite{Ku_Boozer_2009} and near-axis theory \cite{landreman2022mappingQS} QA designs seemed to be limited to either large aspect ratio or low number of field periods.
However, recent work discovered the possibility to analytically perturb low aspect ratio ($A < 2.5$) axisymmetric equilibria to obtain QA configurations with a large range of field period numbers, both in vacuum \cite{plunk_helander_2018} and finite beta \cite{Plunk2020}. 
A recent study by Henneberg and Plunk \cite{HennebergPlunk2024} found that the configurations presented in \cite{Plunk2020} are realizable with only one simple type of QA (``banana") coil in addition to typical tokamak coils.

\begin{figure*} 
    \centering
    \includegraphics[width=\textwidth]{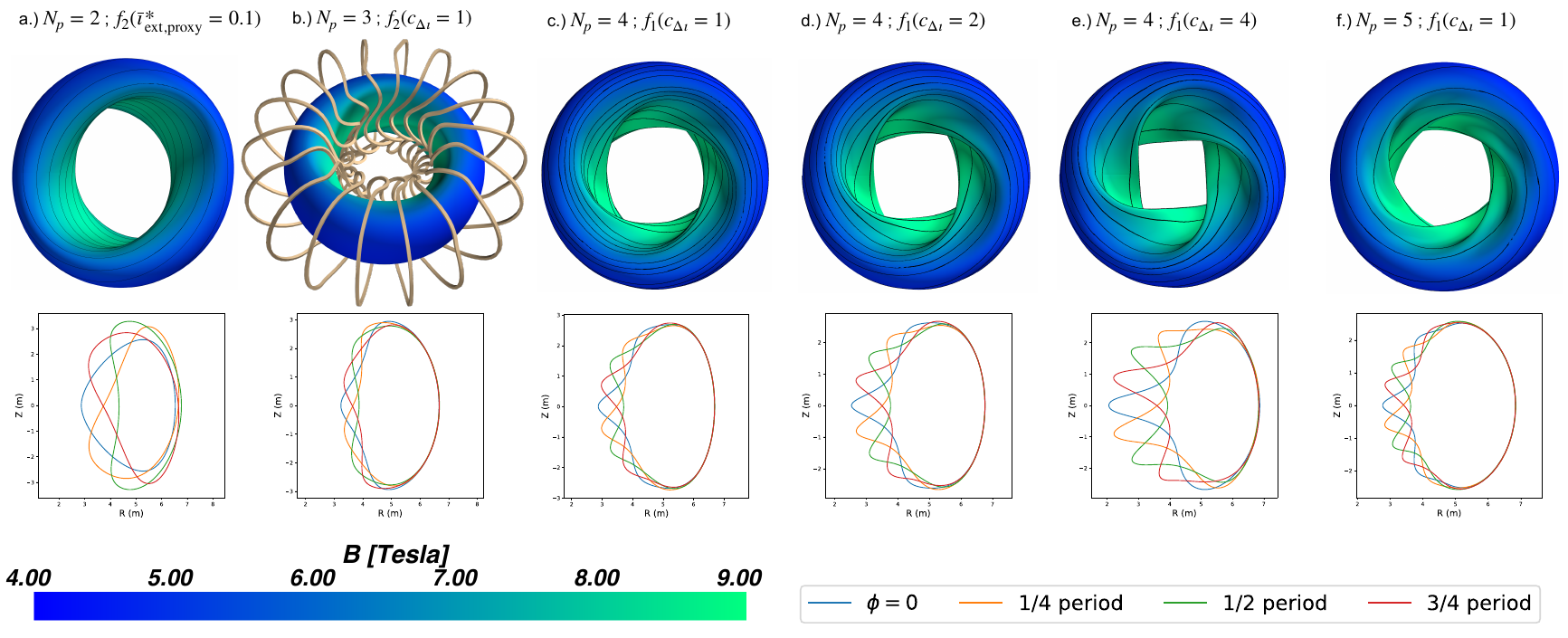}
    \caption{The optimized configurations of different perturbation strengths and field periods. Top plots show the plasma boundary with the field strength and field lines overlaid. Bottom plots show cross-sections within one field period. Optimized modular coils are shown for $N_P = 3$ in (b), and were plotted with the help of CoilPy \cite{coilpy}. The finite coil thickness is for illustrative purposed only and poloidal field coils were omitted for visual clarity.}
    \label{fig:target-demonstration}
\end{figure*}

Here, we numerically optimize and explore this new class of compact QA-stellarators for the first time.
This type of stellarator is difficult to achieve numerically due to its proximity to axisymmetric solutions which are perfectly quasi-axisymmetric.
As we will show, our new target prevents tokamak-like solutions even if we initialize the optimization with an axisymmetric (tokamak) boundary. 
We present improved QA equilibria up to five field periods.
We also demonstrate the possibility of including additional physics targets, e.g. self-consistent bootstrap current and Mercier stability.

\textit{Methods}.\textemdash 
The numerical optimization is performed by minimizing the general total objective
\begin{equation}\label{eq:least-squares-sum}
    f(\mathbf{x}) = \sum_i w_i\: (f_i(\mathbf{x}) - f_i^*)^2.
\end{equation}
Here the $f_i$ represent our physics targets and constraints, $f_i^*$ their desired target value, $w_i$ their weights, and $\mathbf{x}$ represents the free parameters. 
Posing the optimization problem in the form of scalarization as done here has shown to work well in the past \cite{LandremanPaul2022,Henneberg2019}. 
The weights are chosen such that the contributions in Eq. \ref{eq:least-squares-sum} are of similar magnitude for the initial condition and sufficient to satisfy the constraints in the final solution.

We perform the optimization with the SIMSOPT code \cite{landreman2021simsopt, simsoptRelease} using as independent degrees of freedom the Fourier amplitudes of the plasma boundary, the enclosed toroidal magnetic flux, and in two cases additionally the toroidal plasma current density profile.
These are required as inputs for the finite-beta ideal MHD equilibria which are calculated with the VMEC code \cite{HirshmanWhitson1983}.
The nonlinear least squares problem is solved with scipy's default trust region algorithm and gradients are computed with finite differences.
The plasma boundary is represented as a double Fourier series
\begin{align}
    R(\vartheta,\phi) &= \sum_{m=0}^{M}\sum_{n=-N}^{N} R_{m,n} \cos{(m\vartheta-N_{P}n\phi)}, \label{eq:R-fourier} \\
    Z(\vartheta,\phi) &= \sum_{m=0}^{M}\sum_{n=-N}^{N} Z_{m,n} \sin{(m\vartheta-N_{P}n\phi)}, \label{eq:Z-fourier}
\end{align}
where $\phi$ is the geometric toroidal angle and $\vartheta$ a poloidal parameter. The number of boundary modes that are included in the optimization are increased iteratively \cite{LandremanPaul2022}. 
The optimizations of $N_P > 2$ are performed in three steps where $R_{m,n}$ and $Z_{m,n}$ with $m,|n| \leq 4 + j$ are included in step $j = 0,1,2$. Modes at least up to $m, |n| = 4$ must be included in each step to retain the main features of the analytic starting point which was obtained from the $N_{P} = 4$ finite-beta equilibrium presented in \cite{HennebergPlunk2024}. 
The $N_P = 2$ optimization was performed in five steps starting with $m, |n| \leq 1$ and a circular axisymmetric boundary was used as the starting point.

The chosen temperature and density profiles are \cite{Landreman2022bootstrap}
\begin{equation}\label{eq:pressure-profile}
    n_{\alpha}(s) = n_{\alpha,0} (1 - s^5)\quad \text{and}\quad
    T_{\alpha}(s) = T_{\alpha,0} (1 - s),
\end{equation}
with the on-axis values $n_{e,0} = n_{H,0}$ and $T_{e,0} = T_{H,0}$
\footnote{For the $N_{P} = 2,3$ optimization the on-axis density and temperature values were chosen to be $3 \times 10^{20}\:{\mathrm{m^{-3}}}$ and $14\:{\mathrm{keV}}$ to initialize at $\langle\beta\rangle \sim 5\%$. 
For the $N_{P} = 4,5$ cases the initial values were $2.2 \times 10^{20}\:{\mathrm{m^{-3}}}$ and $10\:{\mathrm{keV}}$, yielding $\langle\beta\rangle \sim 2.5\%$. In these cases $\langle\beta\rangle$ generally increases during the optimizations. These values change slightly when the configurations are scaled to the magnetic field of the ARIES-CS design after the optimization.}.
To optimize for QS alone, the objective takes the form
\begin{equation}
f_{1}(c_{\Delta\iota}) = (A-A^*)^2 + f_{QS} + f_\beta + f_{\iota,\mathrm{ext}}.
\end{equation}
Here $A$ is the aspect ratio and is targeted to remain at the initial value, $A^* = 2.42$.
QS is targeted through the following expression which does not require the calculation of Boozer coordinates and vanishes for perfect QS on each flux surface \cite{LandremanPaul2022},
\begin{eqnarray}\label{eq:QS-target}
f_{QS}=&&\sum_{s_j} w_j \Big\langle\big(\frac{1}{B^3}[(N-\iota M)\mathbf{B}\times \nabla B \cdot \nabla\psi\nonumber\\
&&- (G+NI)\mathbf{B}\cdot\nabla B]\big)^2\Big \rangle.
\end{eqnarray}
Here $2\pi G/\mu_0$ is the poloidal current, $2\pi I/\mu_0$ is the toroidal current \cite{Helander_2014}, $N=0$, and $M=1$. The sum in eq. \ref{eq:QS-target} is over 11 flux surfaces $s = \frac{\psi}{\psi_a}=0,0.1,...,1$ with $\psi_a$ being the flux at the plasma boundary. No weighting is applied ($w_j = 1$).
The third term in the objective constrains the volume averaged plasma beta to stay below $5\%$, implemented through quadratic regularization with $f_\beta = w_\beta \ast \max{[\beta-0.05,0]}^2$.

\begin{figure}
    \centering
    \includegraphics[width=0.45\textwidth]{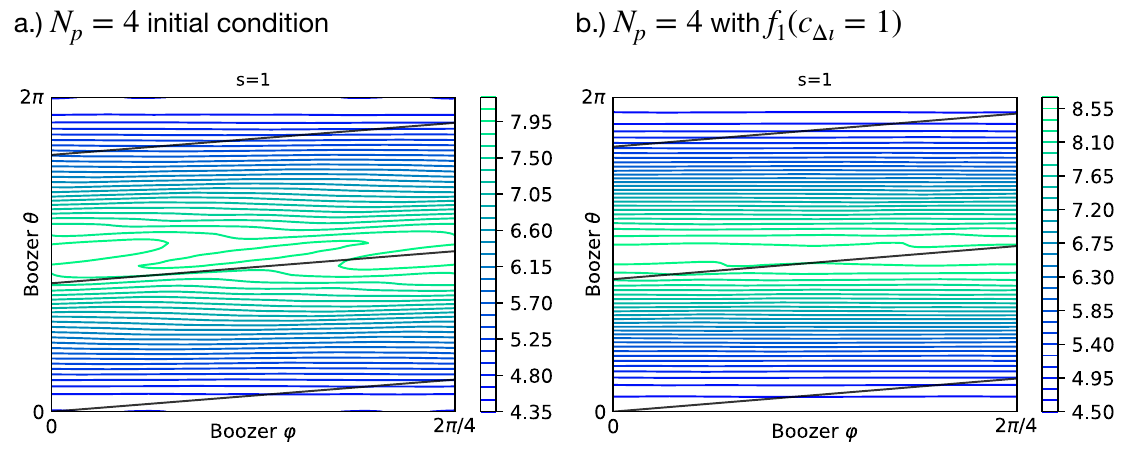}
    \caption{
    The contours of magnetic field strength in Boozer coordinates at the plasma boundary before (a) and after (b) the optimization.
    A section of one field line mapped into one field period is shown as a straight black line.}
    \label{fig:booz-initial-condition-and-QA-optimized}
\end{figure}
\begin{figure}[!b]
    \centering
    \includegraphics[width=0.45\textwidth]{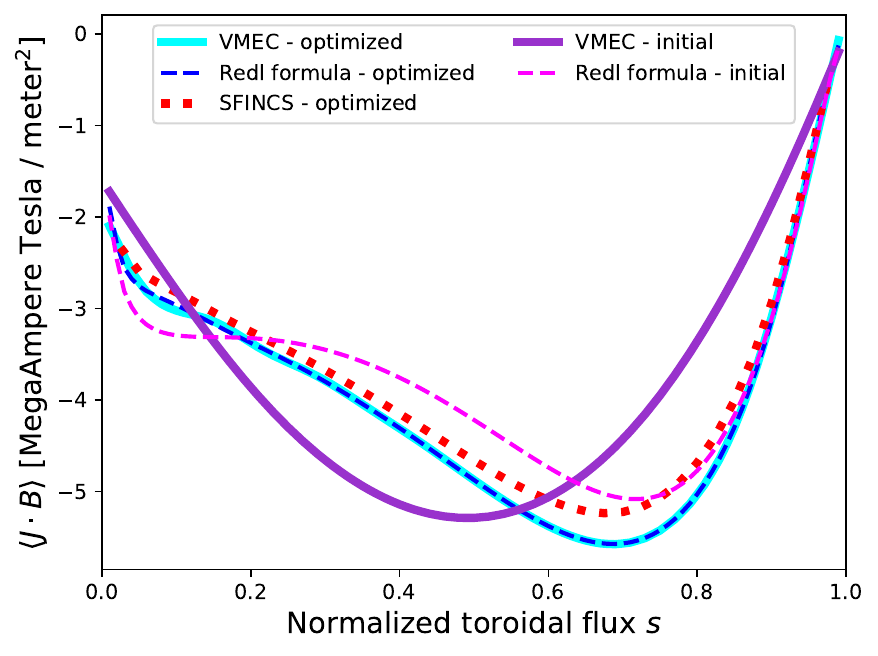}
    \caption{The bootstrap current profile before and after the optimization of $N_p = 3$ for $f_2(c_{\Delta\iota}=1.0)$. 
    The optimized current profile shows self-consistency at $\langle\beta\rangle = 5\%$. 
    With the ARIES-CS volume $V = 444\:\mathrm{m^3}$ and field $\langle B\rangle = 5.86\:\mathrm{T}$ a net toroidal current of $I_\mathrm{tor} = 9.3\:\mathrm{MA}$ is self-generated.}
    \label{fig:bootstrap-profiles}
\end{figure}

The final term in the objective, $f_{\iota,\mathrm{ext}}$, addresses the main challenge in this optimization problem, namely that the QA error decreases monotonically towards the axisymmetric tokamak solution as the strength of the QA perturbation is decreased \cite{Plunk2020, HennebergPlunk2024}.
Unless further constraints are added, the optimizer seeks out axisymmetry for which $f_{QS} = 0$. 
This unique problem of finite-beta QA optimization has been encountered before, where distinctly non-axisymmetric initial conditions were chosen instead \cite{Landreman2022bootstrap}.
Here, a new target has been constructed to find a proxy for the external rotational transform,
i.e. the contribution from 3D shaping. 
For this, an additional equilibrium is being calculated 
at each step in the optimization with the same current profile, pressure profile, and axisymmetric boundary but all non-axisymmetric boundary modes set to zero, i.e. $R_{m,n\neq0} = Z_{m,n\neq0} = 0$. We refer to the mean rotational transform of this equilibrium as $\overline{\iota}_\mathrm{current,proxy}$ since it serves as a proxy for the rotational transform contribution from the current in the QA equilibrium.
By comparing the rotational transform between both equilibria the radially averaged external 
rotational transform can be estimated as $\overline{\iota}_\mathrm{ext,proxy} = \overline{\iota} - \overline{\iota}_\mathrm{current,proxy}$. Even though this target is only a proxy for the true external rotational transform, it works well in practice where it is used as a lower bound to constrain the optimization, i.e. $f_{\iota,\mathrm{ext}} = w_\iota \ast \max{[\overline{\iota}_\mathrm{ext,proxy}^*-\overline{\iota}_\mathrm{ext,proxy},0]}^2$. 
Where applicable we specify the lower bound $\overline{\iota}_\mathrm{ext,proxy}^*$ with respect to the value of the analytic initial condition by setting $\overline{\iota}_\mathrm{ext,proxy}^* = c_{\Delta\iota} \overline{\iota}_\mathrm{ext,proxy}^\mathrm{initial}$, where $\overline{\iota}_\mathrm{ext,proxy}^\mathrm{initial} \sim 0.05$, and varying the external rotational transform factor $c_{\Delta\iota}$.

\begin{figure}[!b]
    \centering
    \includegraphics[width=0.45\textwidth]{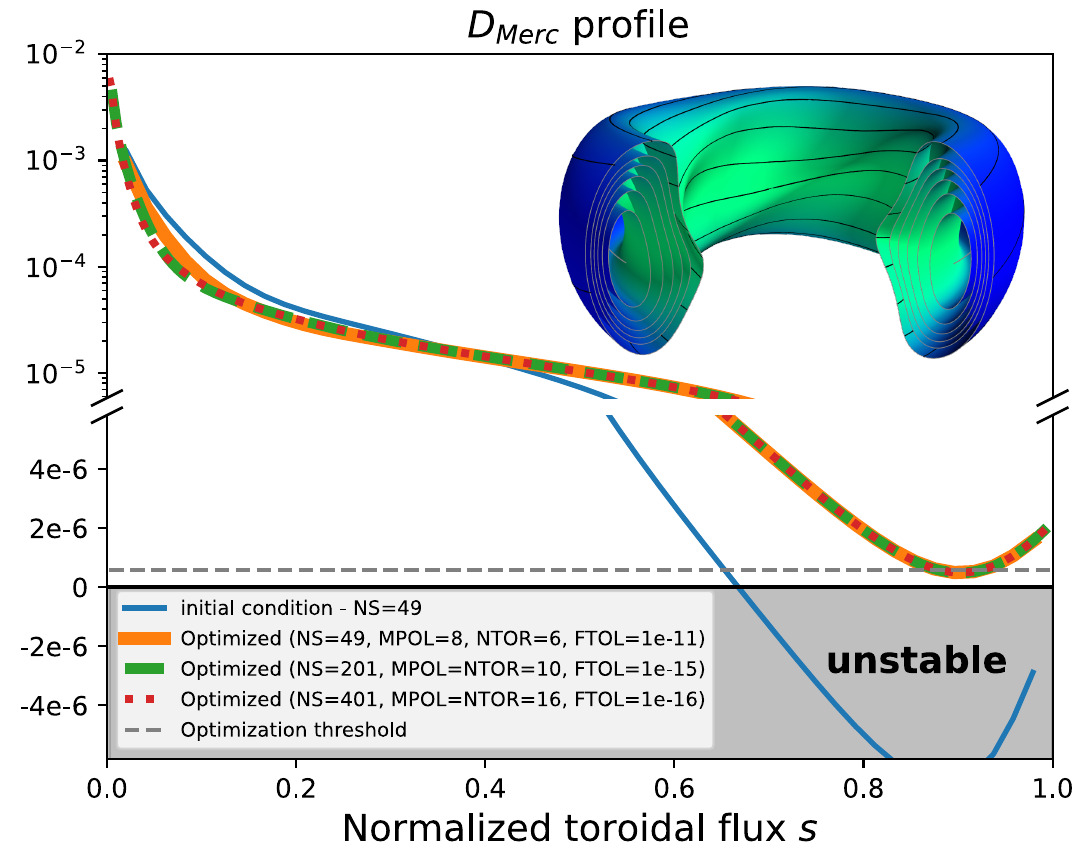}
    \caption{The radial profile of VMEC's Mercier coefficient $D_\mathrm{Merc}$ before and after the optimization of $N_p = 3$ for $f_2(c_{\Delta\iota}=1.0)$.}
    \label{fig:mercier-profile}
\end{figure}

\textit{Results}.\textemdash 
The QA-optimized plasma boundary with four field periods ($N_p = 4$) is shown in Fig. \ref{fig:target-demonstration}(c)-(e) for increasing values of the external rotational transform factor $c_{\Delta\iota}$. 
The effectiveness of our new external rotational transform target $f_{\iota,\mathrm{ext}}$ in setting the desired departure from axisymmetry becomes apparent. 
Another optimized equilibrium with five field periods is shown in Fig. \ref{fig:target-demonstration}(f). 
The optimizations achieve a significant reduction in the QA error as visible in Fig. \ref{fig:booz-initial-condition-and-QA-optimized} which shows the magnetic field strength contours in Boozer coordinates at the plasma boundary before (a) and after the optimization (b). 
The contours in Fig. \ref{fig:booz-initial-condition-and-QA-optimized} (b) appear straighter than those in (a), corresponding to improved QA quality.

In addition to the purely QA optimized cases of four and five field periods we present configurations with two and three field periods optimized for QA, a self-consistent bootstrap current (i.e. assuming no external current drive), and Mercier stability.
The plasma driven bootstrap current is a kinetic effect and is thus not included when solving for the ideal MHD equilibrium. 
Therefore, the equilibrium can predict a net parallel current which does not match the bootstrap current.
To optimize for a self-consistent bootstrap current, we use a recently developed target \cite{Landreman2022bootstrap,landreman_2022_bootstrapDataset} which is based on an analytic tokamak expression by Redl \textit{et al} \cite{Redl2021},
\begin{equation}
    f_{boot} = \frac{\int_0^1 ds [\langle\mathbf{j}\cdot\mathbf{B}\rangle_\mathrm{vmec}- \langle\mathbf{j}\cdot\mathbf{B}\rangle_\mathrm{Redl}]^2}{\int_0^1 ds [\langle\mathbf{j}\cdot\mathbf{B}\rangle_\mathrm{vmec}+ \langle\mathbf{j}\cdot\mathbf{B}\rangle_\mathrm{Redl}]^2}.
\end{equation}
The Mercier criterion ($D_\mathrm{Merc} > 0$) is a sufficient condition for interchange stability and a necessary condition for ballooning stability \cite{GreeneJohnson1962,GreeneJohnson1962extra,Mercier1962,Mercier1964,LandremanJorge2020,bauer2012magnetohydrodynamic}. Although its importance in stellarators remains elusive \cite{freidberg2014,Weller2006,Geiger2004}, it is typically included in designs as a precautionary measure.
Mercier stability was obtained by adding a target $f_\mathrm{Merc}$ similar to $f_\beta$ to set a lower bound for the Mercier parameter $D_\mathrm{Merc}$.
Hence, the general objective now reads
\begin{equation}
    f_2 = f_1 + f_{boot} + f_\mathrm{Merc}.
\end{equation}
For the $N_P = 3$ case we additionally constrained the elongation to stay below $2.0$ and the rotational transform to stay below $0.5$ to avoid the creation of magnetic islands by the most dangerous possible rational surface $\iota = 1/2$. 
Here we define elongation as the ratio between the width at the midplane and the maximum height, both taken at the half period ($\phi = \pi / N_{P}$) toroidal cross-section. These constraints are implemented analogously to $f_\beta$.

\begin{figure}
    \centering
    \includegraphics[width=0.45\textwidth]{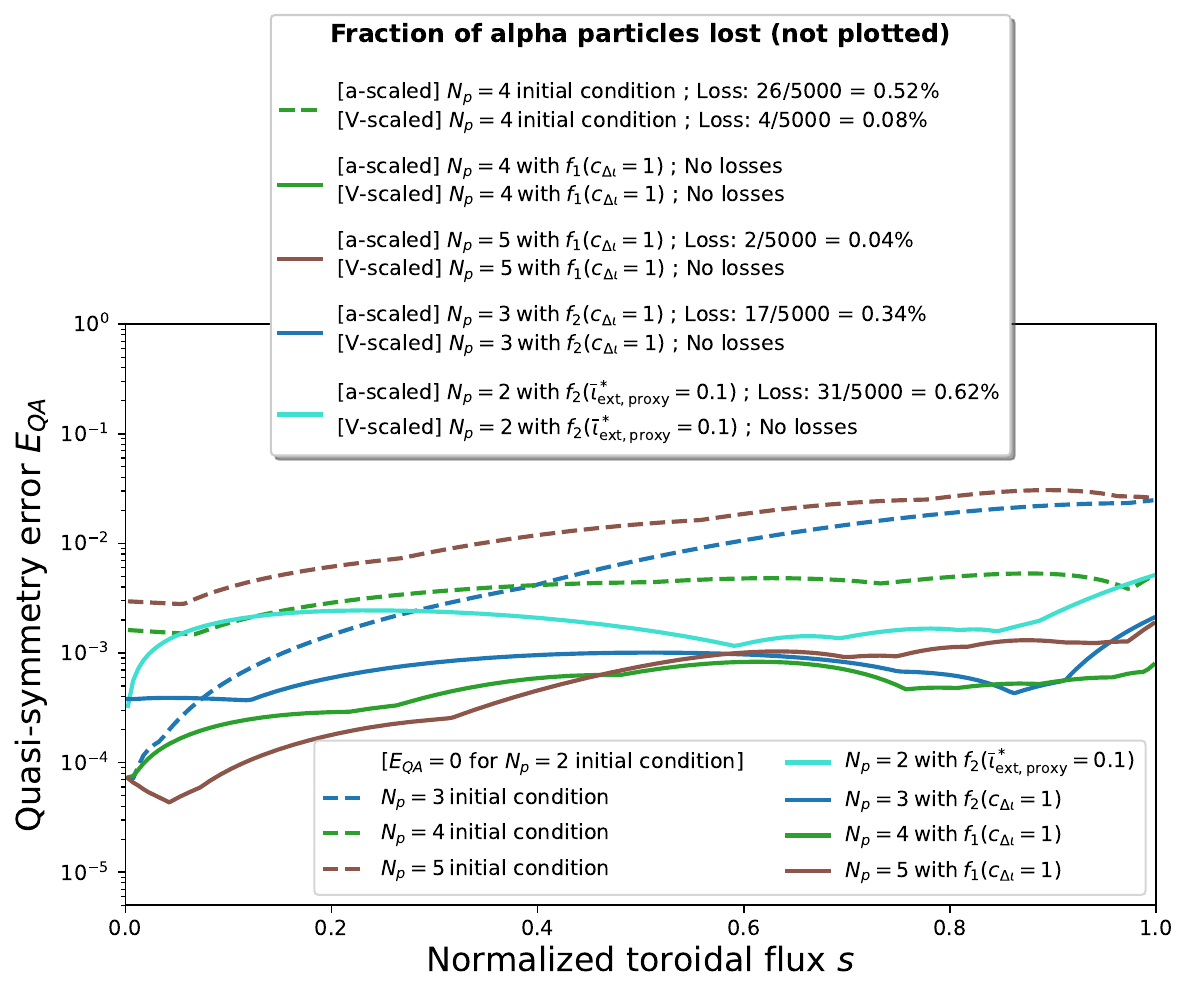}
    \caption{
    The radial profiles of the quasi-symmetry error $E_{QA}$.
    Mode amplitudes were normalized to their mean field $B_{00}$ for this comparison. The legend at the top lists the loss fraction of $3.5\:\mathrm{MeV}$ alpha particles born on the $s=0.25$ flux surface. 
    }
    \label{fig:summaries}
\end{figure}

The optimized plasma boundaries in Fig. \ref{fig:target-demonstration}(a)-(b) show a reduction of the axisymmetric volume from three to two field periods, making the $N_P=2$ case less attractive for a stellarator-tokamak hybrid as proposed in \cite{HennebergPlunk2024} but still an interesting stellarator. 
The bootstrap current profile before and after the optimization is shown for $N_P=3$ in Fig. \ref{fig:bootstrap-profiles}, demonstrating that self-consistency was achieved. 
The Redl prediction for the final result was checked with the general 3D drift-kinetic code SFINCS \cite{SFINCS} and Fig. \ref{fig:bootstrap-profiles} shows good agreement between both \footnotetext[5]{It is worth noting that we have not performed any fixed-point iterations between VMEC and SFINCS which could enhance the agreement further \cite{Landreman2022bootstrap}.}
\cite{Note5, Landreman2022bootstrap,landreman_2022_bootstrapDataset}.
Fig. \ref{fig:target-demonstration}(b) shows that, like the ``banana" coils of Ref. \cite{HennebergPlunk2024}, modular coils are also able to support our equilibria. These were optimized with SIMSOPT using the same targets as in Ref. \cite{HennebergPlunk2024} as well as a penalty on large values of local coil curvature. Like the ``banana" coils they create flux surfaces in vacuum. 
Thus, any additional current, like in the non-bootstrap-consistent cases, must not be driven by transformer action (as usually required for start-up in tokamaks) but could be supplied through established microwave or beam current drive.
The initial and final radial profiles of the Mercier parameter are shown in Fig. \ref{fig:mercier-profile}. As it has been previously shown that VMEC's Mercier parameter is sensitive to the radial resolution \cite{LandremanJorge2020,DESC1}, equilibrium calculations up to $N_s = 401$ radial points were performed. 
We find that Mercier stability is possible which might be surprising since the QA perturbations live on the side of good curvature. 
Fig. \ref{fig:summaries} shows the QA quality over the full radial domain, where $E_{QA} = \max\{|B_{m,n \neq 0}| \} / B_{0,0}$ and the magnetic field is taken to be in Boozer coordinates \cite{LandremanPaul2022}. 
Where applicable, the QA quality improved significantly due to the optimization.

\begin{figure}
    \centering
    \includegraphics[width=0.45\textwidth]{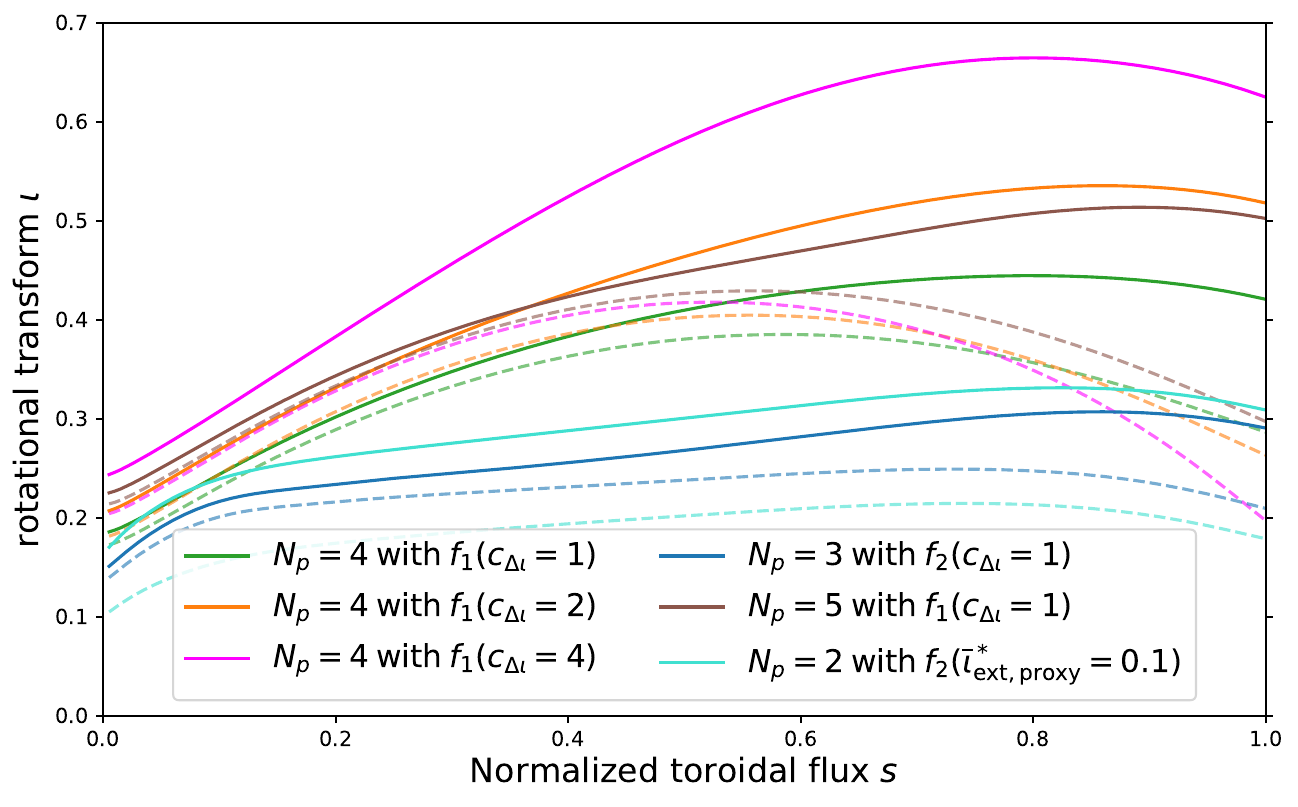}
    \caption{
    The rotational transform profiles of the optimized configurations (solid lines) and their underlying tokamak configurations which are used to evaluate $\overline{\iota}_\mathrm{ext,proxy}$ (dashed lines).
    }
    \label{fig:iota-summary}
\end{figure}

The confinement of $3.5\:\mathrm{MeV}$ alpha particles has been tested with the SIMPLE code \cite{albert_kasilov_kernbichler_2020, Albert2020} where 5000 test particles were assumed to be evenly born on the $s = 0.25$ flux surface and followed for $250\:\mathrm{ms}$. 
To aid comparison with previous results \cite{LandremanPaul2022, Bader2021} all configurations were scaled to the magnetic field and length scale of the ARIES-CS reactor design \cite{ARIES-CS2008}. The top legend in Fig. \ref{fig:summaries} lists the final loss fractions and includes results both for scaling to the same minor radius (a-scaling) as well as to the same volume (V-scaling) as ARIES-CS. Even though the minor radius scaling is suitable for a comparison of the physics performance, it is the scaling to the same volume which compares reactor designs of similar fusion power. 
While the losses of the $N_{P}=4$ initial condition are already low, the significant improvement in QA quality results in no losses for any of the volume-scaled optimized cases. This is also true for the $N_{P} = 2,3$ configurations, even though two additional physics targets were achieved simultaneously. For the minor radius scaling all losses still remain below $1\%$ which is significantly lower than those of any constructed stellarator (see figure 6 in \cite{LandremanPaul2022}).
The rotational transform, shown in Fig. \ref{fig:iota-summary}, is significant even for the bootstrap-consistent cases which assume no external current drive.

\textit{Conclusion}.\textemdash 
It was shown that the new class of compact QA stellarators with a wide range of field periods can be numerically optimized with the help of a new objective function, resulting in equilibria with excellent QA quality while achieving additional physics properties such as MHD stability and a self-consistent bootstrap current.
Since many of these equilibria can be obtained through continuous deformation of a tokamak \cite{Plunk2020,HennebergPlunk2024}, our results further motivate a QA-tokamak hybrid device which could investigate fundamental questions on MHD stability and turbulent transport under continuous departure from axisymmetry and otherwise very similar neoclassical properties. \\

The authors would like to thank Gabriel Plunk for helpful discussions, Istv\' an Cziegler and Per Helander for carefully reading the manuscript, H\aa kan Smith for help with SFINCS, and the SIMSOPT developers for general support. 
The simulations and optimizations were performed on the Cobra (Germany) and Viper (Germany) supercomputers. 
This work has been carried out within the framework of the EUROfusion Consortium, funded by the European Union via the Euratom Research and Training Programme (Grant Agreement No 101052200 — EUROfusion). Views and opinions expressed are however those of the author(s) only and do not necessarily reflect those of the European Union or the European Commission. Neither the European Union nor the European Commission can be held responsible for them.
This work was supported by the Engineering and Physical Sciences Research Council grants [EP/S022430/1].


\providecommand{\noopsort}[1]{}\providecommand{\singleletter}[1]{#1}%

\end{document}